\documentclass[aps,prb,twocolumn,groupedaddress,showpacs]{revtex4}

\usepackage{graphicx}

\begin{document}

\title{Charge ordering and self-assembled nanostructures in
a fcc Coulomb lattice gas}

\author{Khang Hoang,$^{1}$ Keyur Desai,$^{2}$ and S. D.
Mahanti$^{1}$} \email[Author to whom correspondence should be
addressed. Electronic address: ]{mahanti@pa.msu.edu}
%\thanks{}
%\altaffiliation%{}
\affiliation {$^{1}$Department of Physics and Astronomy, Michigan
State University, East Lansing, Michigan 48824, USA
\\ $^{2}$Department of Electrical and Computer Engineering,
Michigan State University, East Lansing, Michigan 48824, USA}

%\date{\today}

\begin{abstract}
The compositional ordering of Ag, Pb, Sb, Te ions in
(AgSbTe$_{2}$)$_{x}$(PbTe)$_{2(1-x)}$ systems possessing a NaCl
structure is studied using a Coulomb lattice gas (CLG) model on a
face-centered cubic (fcc) lattice and Monte Carlo simulations. Our
results show different possible microstructural orderings. Ordered
superlattice structures formed out of AgSbTe$_{2}$ layers separated
by Pb$_{2}$Te$_{2}$ layers are observed for a large range of $x$
values. For $x=0.5$, we see an array of tubular structures formed by
AgSbTe$_{2}$ and Pb$_{2}$Te$_{2}$ blocks. For $x=1$, AgSbTe$_{2}$
has a body-centered tetragonal (bct) structure which is in agreement
with previous Monte Carlo simulation results for restricted
primitive model (RPM) at closed packed density. The phase diagram of
this frustrated CLG system is discussed.
\end{abstract}

%\keywords{[{\it This article has been published in} Phys. Rev. B {\bf 72}, 2005]}

\maketitle

\section{\label{sec:1}Introduction}

Lattice gas with long-range Coulomb interaction has attracted
considerable interest over the past 10 years. Two types of
long-range models have been studied. One where the interaction
between the charges $\propto 1/r$ (Coulomb lattice gas, or CLG), and
the other where the interaction $\propto \ln r$ (lattice Coulomb
gas, or LCG). Studies of various models of one\cite{king1, king2,
king3, levashov}- and two\cite{levashov, teitel1,
teitel2}-dimensional CLG and LCG using different methods have shown
the existence of multiple phase transitions, complexity in phase
diagrams and their practical applications to real materials, e.g.,
KCu$_{7-x}$S$_{4},$\cite{king1, king2, king3}
Ni$_{1-x}$Al$_{x}$(OH)$_{2}$(CO$_{3}$)$_{x/2}$.$y$H$_{2}$O,...
.\cite{levashov} In three-dimensional CLG on a simple cubic (sc)
lattice, several works have been done using either theoretical
calculations (mean-field approximation\cite{stell} and Pad\'{e}
expansion\cite{walker}) or Monte Carlo (MC) simulations.\cite{stell,
van} However, to the best of our knowledge, there are no extensive
studies of CLG on a fcc lattice excepting when all the lattice sites
are occupied by either a positive or a negative charge.\cite{bresme}
It is well known that fcc lattice involves
frustration.\cite{toulouse} Since the role played by frustration in
the nature of phase transition in Ising-type systems (on triangular
or fcc lattice) has been of great interest in statistical
physics,\cite{wannier, alexander, heilmann, phani, kammerer} it is
of equal interest to see what role frustration effects play in
long-range Coulomb systems.

From materials perspective, a quaternary compound
Ag$_{n}$Pb$_{m}$Sb$_{n}$Te$_{m+2n}$ has recently emerged as a
material for potential use in efficient thermoelectric power
generation. It has been found that for low concentrations of Ag, Sb
and when doped appropriately, this system exhibits a high
thermoelectric figure of merit $ZT$ of $\approx$2.2 at 800
K.\cite{hsu} It is one of the best known bulk thermoelectrics at
high temperatures. Quantitative understanding of its properties
requires understanding of atomic structure. Experimental
data\cite{hsu, eric} suggest that this system belongs to an entire
family of compounds, which are compositionally complex yet they
possess the simple cubic NaCl structure on average, but the detailed
ordering of Ag, Pb and Sb ions is not clear. However, as pointed out
by Bilc {\it et al.},\cite{bilc} the electronic structure of these
compounds depends sensitively on the nature of structural
arrangements of Ag and Sb ions. Hence a simple but accurate
theoretical model is necessary to understand and predict the
ordering of the ions in these systems. In this paper we present a
simple ionic model of Ag$_{n}$Pb$_{m}$Sb$_{n}$Te$_{m+2n}$ that
explicitly includes the long-range Coulomb interaction and in which
the ions are located at the sites of a fcc lattice. As will be shown
in the next section, this problem maps onto a spin-$1$ Ising model
on a fcc lattice with long-range antiferromagnetic interaction. We
present details of the model in Sec. \ref{sec:2}. In Sec.
\ref{sec:3} we discuss our Monte Carlo simulation results including
a full phase diagram in the $x-T$ plane. The summary is presented in
Sec. \ref{sec:4}.

\section{\label{sec:2}Model}
%\subsection{}
%\subsubsection{}
We use a model where the minimization of electrostatic interaction
between different ions in the compounds can lead to the
compositional ordering that exists in the system.\cite{hsu} The
total electrostatic energy is then expressed as
\begin{equation}\label{en1}
    E=\frac{e^{2}}{2}\sum_{l\tau \neq l^{'}\tau^{'}}\frac{Q_{l\tau}Q_{l^{'}\tau^{'}}}{\epsilon
    \mid\mathbf{R}_{l\tau}-\mathbf{R}_{l^{'}\tau^{'}}\mid}\, ,
\end{equation}
where $\epsilon$ is the static dielectric constant,
$\mathbf{R}_{l\tau}$ and $Q_{l\tau}$ are, respectively, the position
and charge of an atom at site $\tau$ of cell $l$. This model has
been successfully applied to cubic perovskite alloys. \cite{van}
Here we consider supercells of the NaCl-type structure made of two
interpenetrating fcc lattices with possible mixtures of different
atomic species on Na sites, i.e., $\tau=\{$Na(Ag,Sb,Pb), Cl(Te)$\}$,
with periodic boundary condition. Alloying occurs on the Na
sublattice. In a simple ionic model of
Ag$_{n}$Pb$_{m}$Sb$_{n}$Te$_{m+2n}$, we can assume the Pb ion to be
$2^{+}$, Te ion to be $2^{-}$, Ag ion to be $1^{+}$, and Sb ion to
be $3^{+}$, i.e.,
$Q_{l\tau}=\{Q_{l,\mathrm{Na}};Q_{l,\mathrm{Cl}}\}=\{+1,+3,+2;-2\}$,
where $Q_{l,\mathrm{Cl}}=q_{\mathrm{Cl}}=-2$ is independent of $l$.
Focusing on the Na sublattice sites where ordering occurs, we write
$Q_{l\tau,\mathrm{Na}}=q_{\mathrm{Na}}+\Delta q_{l}$, where
$q_{\mathrm{Na}}=+2$ and $\Delta q_{l}=\{-1,+1,0\}$. Substituting
the expression for $Q_{l,\mathrm{Na}}$ into Eq. (1), we can write
$E=E_{0}+E_{1}+E_{2}$, where the subscripts refer to the number of
powers of $\Delta q_{l}$ appearing in that term. Then $E_{0}$ is
just a constant, it is the energy of an ideal PbTe lattice; $E_{1}$
vanishes due to charge neutrality. The only term which depends on
the charge configuration is $E_{2}$; it is given by
\begin{equation}\label{en2}
    E_{2}=\frac{e^{2}}{2\epsilon a}\sum_{l \neq l^{'}}\frac{\Delta q_{l}\Delta q_{l^{'}}}{
        \mid\mathbf{l}-\mathbf{l^{'}}\mid}
        \equiv \frac{J}{2} \sum_{l \neq l^{'}}\frac{s_{l}s_{l^{'}}}{\mid\mathbf{l}-\mathbf{l^{'}}\mid}
        \, ,
\end{equation}
where ion positions are measured in unit of the fcc lattice constant
$a$; $l$ and $l^{'}$ run over the $N$ sites of the Na-sublattice of
NaCl structure. Thus if we start from a PbTe lattice as a reference
system and replace two Pb ions by one Ag ion and one Sb ion, we map
the system unto an effective CLG with effective charges -1, +1 (of
equal amount) and 0; this implies a constraint, $\sum_{l}\Delta
q_{l}=0$. The model therefore maps onto a spin-$1$ Ising model
($s_{l}=0,\pm 1$) with long-range antiferromagnetic interaction. The
short-range version of this model [nearest- (n.n.) and
next-nearest-neighbor (n.n.n.) interaction],\cite{heilmann, phani,
kammerer} a generalization of this model by adding a n.n
ferromagnetic interaction\cite{grousson} and a continuum version of
this model that takes into account the finite size of the charged
particles (RPM or charged hard sphere model)\cite{stellwu, orkoulas,
pana, bresme, kobelev, ciach} have been investigated. Comparison
with these works will be made in Sec. \ref{sec:3}.

Because of the attraction between +1 and -1 charges, the Ag and Sb
ions tend to come together and form clusters or some sort of ordered
structures depending on the temperature at which these compounds are
synthesized and the annealing scheme. The ordering may be quite
complex compared to the one on a simple cubic lattice because of the
frustration associated with spins on a fcc lattice and
antiferromagnetic interaction (in the Ising model). In our
calculations of Ag$_{n}$Pb$_{m}$Sb$_{n}$Te$_{m+2n}$, an equivalent
formula, (AgSbTe$_{2}$)$_{x}$(PbTe)$_{2(1-x)}$, is used; where
$x=2n/(m+2n)=1/N\sum_{l}|\Delta q_{l}|\equiv 1/N\sum_{l}s_{l}^{2}$
($0\leq x \leq 1$), is the concentration of Ag and Sb in the Pb
sublattice.

\section{\label{sec:3}Simulation Results}
To study the thermodynamic properties and microstructural ordering
of the system, we have done canonical ensemble Monte Carlo
simulations following the usual Metropolis criterion \cite{metro}
using the energy given by Eq. (\ref{en2}), i.e., particles interact
via site-exclusive (multiple occupancy forbidden) Coulomb
interaction. In the Ising model problem, this corresponds to a fixed
magnetization simulation. We used Ewald summation \cite{ewald} to
handle this long-range interaction employing a very fast lookup
table scheme using Hoshen-Kopelman algorithm.\cite{hk} This model is
parameter-free in the sense that $J=e^{2}/\epsilon a$ defines a
characteristic energy.

A simulation for a fixed concentration $x$ starts at a high
temperature with an initial random configuration followed by gradual
cooling. For each temperature $T$, we use $2\times 10^{4}$ sweeps
(MC steps per lattice site) to get thermal equilibration followed by
$10^{5}$ sweeps for averaging. Particles move either via hopping to
empty sites or via exchange mechanism. The equilibrium configuration
at a given temperature $T$ is used as the initial configuration for
a study at a nearby temperature. We monitored different
thermodynamic quantities and look at the microstructures. The data
presented below were obtained with system size $L=8$ (i.e., $8$ fcc
cells in one direction, 2048 lattice sites in total) with periodic
boundaries.

\begin{figure}[h]
%\hspace{-0.5in}
 %\vspace{0.8in}
 \includegraphics[height=2.5in,width=3.5in,angle=0]{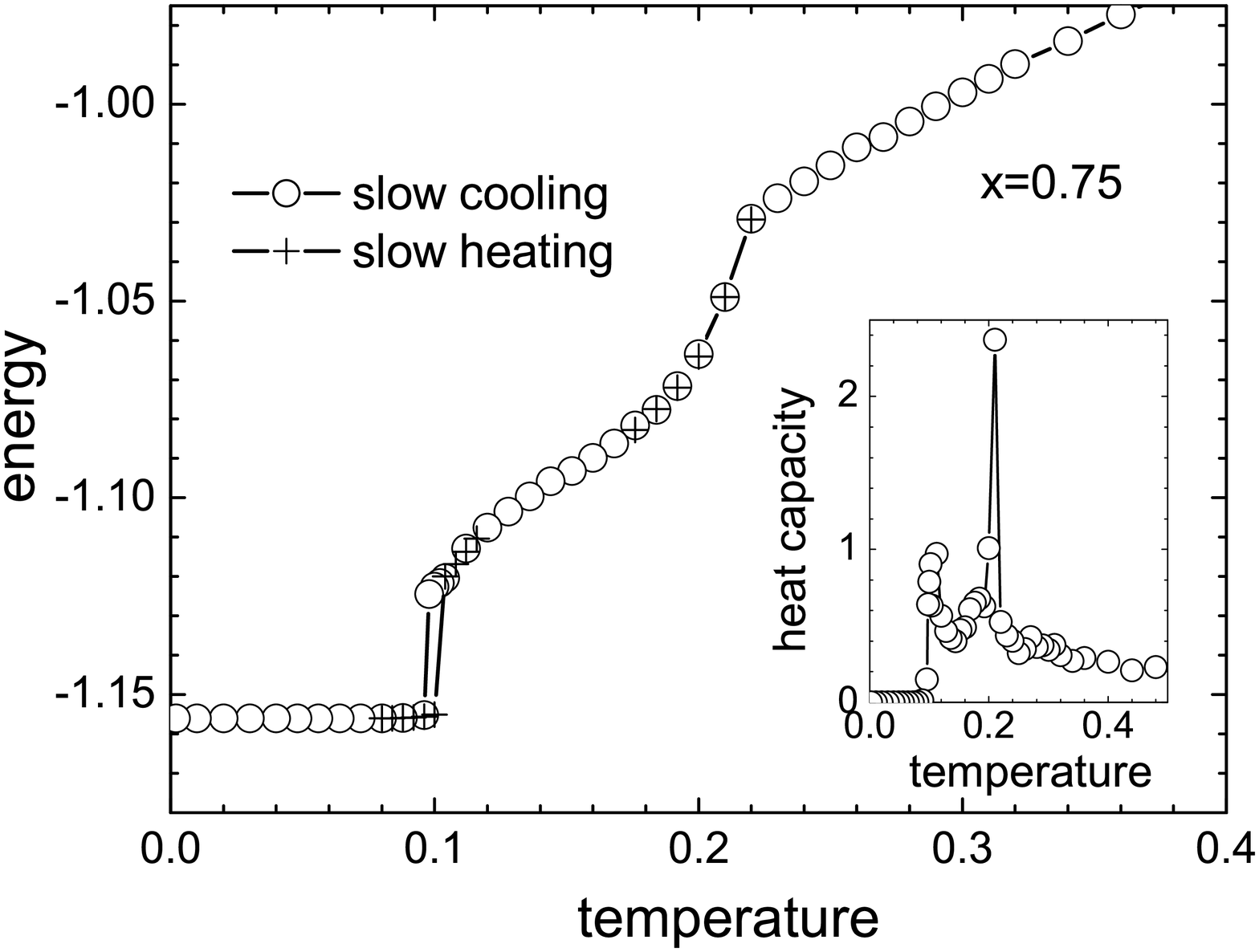}
 \vspace{-0.2in}
 \caption{\label{enerx75} Energy and heat capacity per particle versus temperature for $x=0.75$.
 Phase transitions occur at $T=0.106$ and $0.21$ which
 are first-order and second-order transitions, respectively.
 There is hysteresis associated with the low $T$ transition.}
 \end{figure}
\begin{figure}[h]
%\hspace{-0.5in}
 %\vspace{0.8in}
 \includegraphics[height=2.5in,width=3.5in,angle=0]{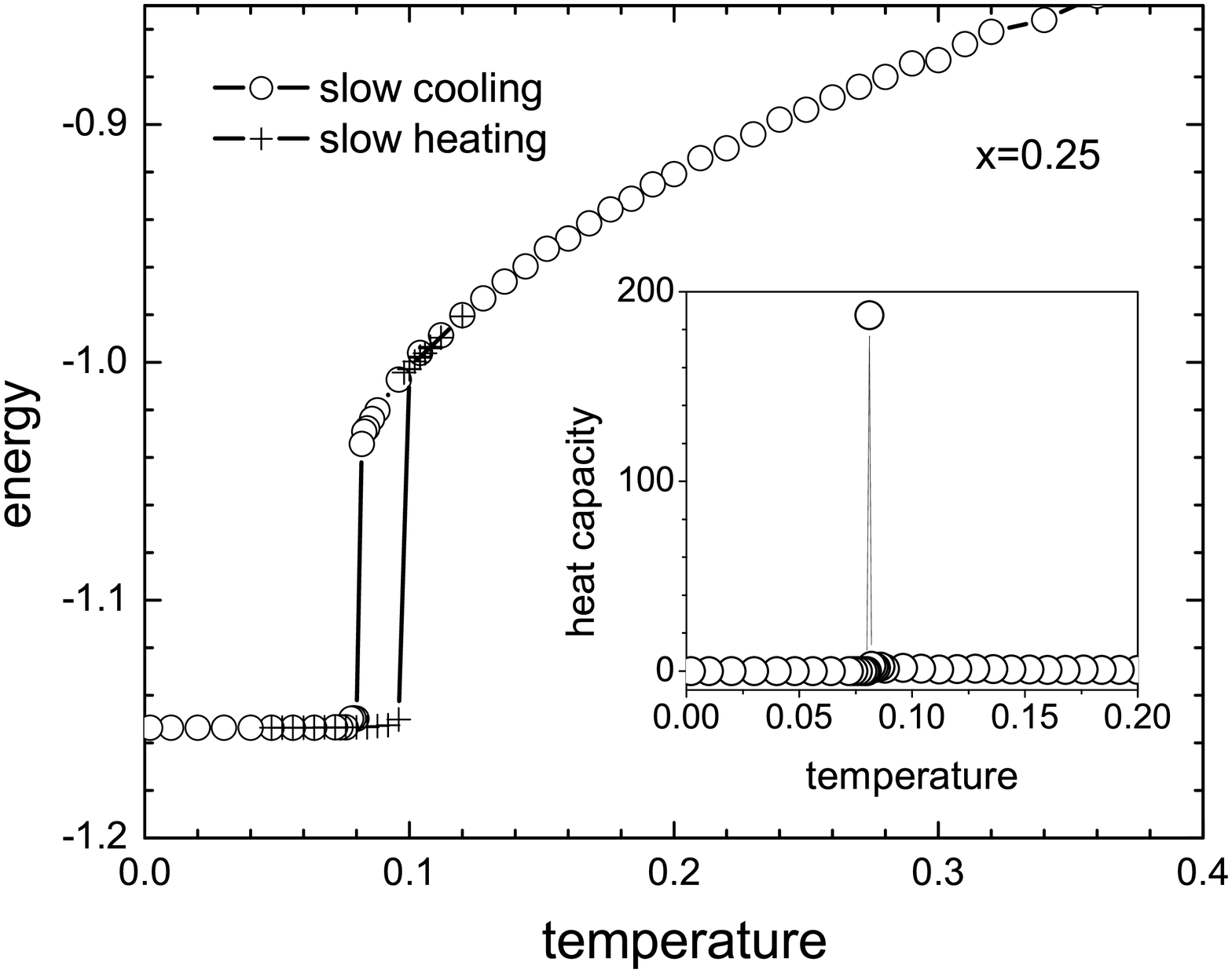}
 \vspace{-0.2in}
 \caption{\label{enerx25} Energy and heat capacity per particle versus temperature for $x=0.25$.
 Phase transition occurs at $T=0.08$ which
 is first-order. There is hysteresis associated with this transition.}
 \end{figure}

Figure \ref{enerx75} shows the energy and heat capacity (obtained
using energy fluctuation) for $x=0.75$ where the two energy curves
correspond to slow cooling and slow heating. We see evidence of two
phase transitions, one at $T=0.106$ and the other at $T=0.21$. The
heat capacity curve shows peaks at the above two $T$ values. The
transition at higher $T$ is continuous and indicates a lattice
gas-liquidlike phase transition. There is no apparent hysteresis
associated with this transition. The low $T$ transition, on the
other hand, appears to be first-order. There is an energy
discontinuity and there is hysteresis, albeit small, associated with
this transition. For $x \leq 0.5$ we see only one transition which
is first-order (Fig. \ref{enerx25}).
 This suggests that with decreasing $x$ the system changes from
undergoing 2 to 1 phase transition. As $x$ decreases from $0.875$,
the high $T$ continuous and low $T$ first-order phase transitions
approach each other and the two transitions merge at $x\approx 0.5$.
As $x$ increases from $0.875$, the high $T$ continuous transition
changes to first-order. At $x=1$, the transition is first-order in
agreement with previous simulation results.\cite{bresme} We also
monitored the structure factor $S(\mathbf{q})$ for different
$\mathbf{q}$ values. For several $\mathbf{q}$ values, we find that
$S(\mathbf{q})$ changes discontinuously at the first-order
transition and smoothly at a continuous transition.

\begin{figure}[h]
%\hspace{-0.25in}
%\vspace{0.8in}
\includegraphics[height=2.6in,width=3.5in,angle=0]{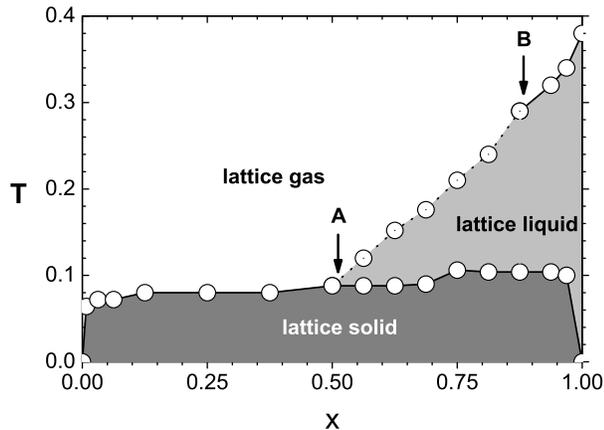}
\vspace{-0.25in}
 \caption{\label{phases} Concentration ($x$) versus temperature ($T$) phase diagram constructed from the loci of the heat capacity maxima.
 There are first (solid lines) and second (dotted line) order transitions with two possible tricritical
 points ($x_{t}$, $T_{t}$): A at $\approx(0.500, 0.088)$ and B at $\approx(0.875, 0.290)$.}
 \end{figure}
To construct the total phase diagram, we have studied energy, heat
capacity and structure factor as functions of temperature for a
series 15 values of the concentration $x$. Figure \ref{phases} shows
the phase diagram constructed from the loci of specific heat maxima.
The lattice gas-lattice solid and lattice liquid-lattice solid
transitions are first-order. In the limit $x=0$, the compound is
simply PbTe, the transition occurs at $T=0$ since there are no
charged particles (effective charges of Pb and Te are 0). For $x=1$,
the compound is AgSbTe$_{2}$. Our simulations show a strongly
first-order transition at $T=0.38$ and no other transition with
decreasing $T$. This strong first-order transition is softened by
introducing defects into the system (by decreasing $x$ from 1). The
hysteresis associated with this transition becomes smaller with
decreasing $x$ from 1 and disappears at $x \approx 0.875$ showing a
changeover from a first- to a second-order transition. Therefore, we
have two possible tricritical points ($x_{t}$, $T_{t}$): A at
$\approx(0.500, 0.088)$ and B at $\approx(0.875, 0.290)$. More
accurate results on the tricritical points would require further
careful large-scale simulations for more number of $x$ values, and
perhaps much larger systems.

We would now like to compare our results with those of previous
simulations carried out for lattice RPM. In this model, there is a
parameter $\xi=\sigma/a$, where $\sigma$ is the hard sphere diameter
of the charged particles. For $\xi =1$ which is comparable to our
model, Dickman and Stell\cite{stell} and Panagiotopoulos and
Kumar\cite{pana} have a phase diagram for a sc lattice that is
similar to ours. They found a tricritical point at ($x_{t}$,
$T_{t}$) $\simeq$ ($0.4$, $0.14$) (Ref. 7) and ($0.48\pm 0.02$,
$0.15\pm 0.01$).\cite{pana} It appears that $x_{t}$ values for sc
and fcc lattices are quite close whereas the $T_{t}$ values for the
fcc lattice is about a factor of 0.6 smaller, perhaps due to
frustration. As regards the second tricritical point (B), it is
unique to the fcc lattice. Dickman and Stell\cite{stell} found a
high $T$ continuous phase transition ($\lambda$-transition) in a
simple cubic lattice as $x$ increased from $0.4$ to $0.82$. The
observation of the high $T$ first-order transition in our
simulations is similar to the one seen in fully-frustrated n.n. and
n.n.n. Ising model ($x=1$) seen by Phani {\it et al.}\cite{phani}

\begin{figure}[h]
%\hspace{-0.5in}
%\vspace{0.8in}
 \includegraphics[height=2.5in,width=3.5in,angle=0]{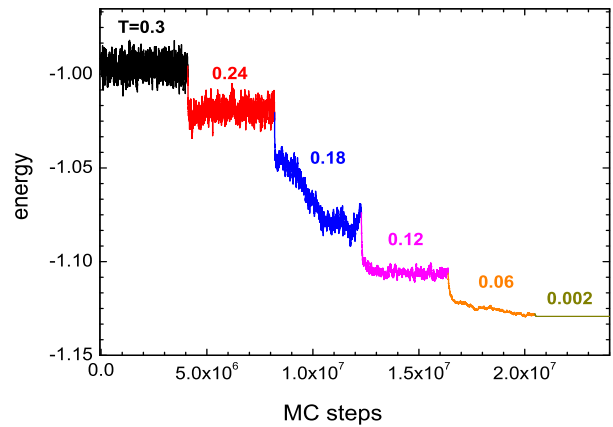}
\vspace{-0.2in}
 \caption{\label{EvsMCsteps} (Color online) Energy as a function of Monte Carlo steps for $x=0.75$.
  The system is quenched from $T=0.3$ to $0.24$, then from $T=0.24$ to $0.18$,
  and so on.}
 \end{figure}
In Fig. \ref{EvsMCsteps}, we plot the energy of a system with
$x=0.75$ as a function of Monte Carlo steps as we quench the system
from $T=0.3$ to $0.002$ through several intermediate values of $T$.
We use more than $4\times 10^{6}$ moves for each $T$ without
discarding any step for thermal equilibration. The final
configuration at a given $T$ is used as the initial configuration
for the next $T$. The fluctuation is large at high $T$ and getting
smaller with lowering $T$. From one $T$ to another, it takes some
time ($\approx 10^{3}$ steps) for the system to equilibrate. As one
crosses the transition region, i.e., from $T=0.24$ to $0.18$ or from
$0.12$ to $0.06$, the result shows the existence of possible local
minima in energy where the system is in metastable states and then
goes to a stable state with lower energy. More details on quenching
studies will be reported in another paper.\cite{hoang}

\begin{figure}[h]
%\hspace{-0.5in}
\vspace{-0.1in}
 \includegraphics[height=2.5in,width=2.5in,angle=0]{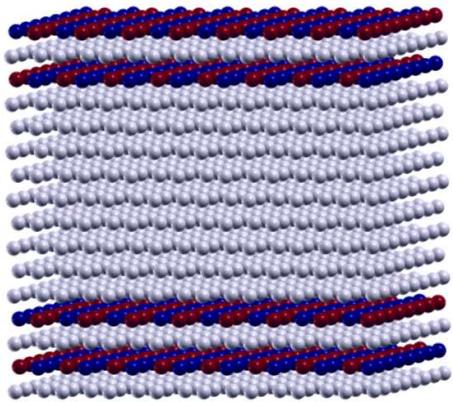}
 \vspace{-0.2in}
 \caption{\label{configx25} (Color online) A low temperature configuration for $x=0.25$ [created using XCrySDen (Ref. 30)].
 Dark layers are for Ag/Sb, grey layers are for Pb; Te sublattice is not
 shown. This typical configuration showing a ordered superlattice structure formed out
 of AgSbTe$_{2}$ layers separated by Pb$_{2}$Te$_{2}$ layers.}
 \end{figure}
A typical low temperature structure of
(AgSbTe$_{2}$)$_{x}$(PbTe)$_{2(1-x)}$ is a self-assembled
nanostructure with layers of AgSbTe$_{2}$ arranged in a particular
fashion in the PbTe bulk as shown in Fig. \ref{configx25} for the
case $x=0.25$. Four layers of AgSbTe$_{2}$ are separated from one
another by four layers of Pb$_{2}$Te$_{2}$. This domain is again
separated by a purely PbTe domain formed by eight other layers of
Pb$_{2}$Te$_{2}$. Along the $z$-direction (perpendicular to the
layers), positive charge and negative charge arrange consecutively.
This indicates a three-dimensional long-range order which is clearly
a result of the long-range Coulomb interaction. Experimentally,
high-resolution transmission electron microscopy (TEM) images
indicate inhomogeneities in the microstructure of the materials,
showing nano-domains of a Ag-Sb-rich phase embedded in a PbTe
matrix\cite{hsu, eric} which appears to be consistent with our
results. Also electron diffraction measurements show clear
experimental evidence of long-range ordering of Ag and Sb ions in
AgPb$_{m}$SbTe$_{m+2}$ (m=18).\cite{bilc}

\begin{figure}[h]
%\hspace{-0.5in}
\vspace{0.1in}
 \includegraphics[height=2.2in,width=2.2in,angle=0]{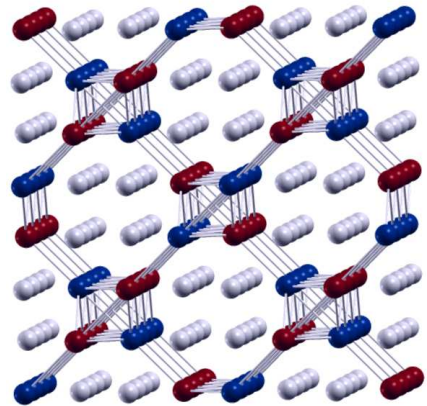}
 %\vspace{-0.04in}
 \caption{\label{configx50} (Color online) A low temperature configuration for $x=0.5$ [created using XCrySDen (Ref. 30)].
 Connected balls are for Ag/Sb, unconnected balls are for Pb; Te sub-lattice is not shown.
 Checkerboard pattern formed by AgSbTe$_{2}$ and Pb$_{2}$Te$_{2}$ blocks.}
 \end{figure}
In addition to the layered superlattice structures seen for several
$x$ values, we have also discovered a very interesting structure at
$x=0.5$, in which an array of tubes of AgSbTe$_{2}$ and
Pb$_{2}$Te$_{2}$ are arranged in a checker board pattern (Fig.
\ref{configx50}). We find that this structure has the same energy as
the layered structure consisting of alternate layers of AgSbTe$_{2}$
and Pb$_{2}$Te$_{2}$ (energy per particle $E=-1.157278$). For $x=1$,
i.e., AgSbTe$_{2}$, the only ordered structure is body-centered
tetragonal (bct) structure with a $c$-parameter which is double that
of the NaCl subcell, belonging to space group $I\overline{4}m2$. The
unit cell of this structure has eight ions, with every ion being
surrounded by eight ions of opposite charge and four of the same
charge (Te sublattice is not included here). This structure is
equivalent to the type-III antiferromagnetic
structure\cite{andersonsmart} which has been found in n.n. and
n.n.n. Ising model by Phani {\it et al.}\cite{phani} It has also
been seen in the RPM by Bresme {\it et al.}\cite{bresme} in Monte
Carlo simulations and by Ciach and Stell\cite{ciach} within a
field-theoretic approach.

The comparison we made with a system of size $L=4$ shows no
appreciable change in the results for the first-order transition
except the fact that we did not see any hysteresis with $L=4$. The
energy at a given concentration differs by $0.1\%-0.5\%$ from that
obtained for $L=8$. However, for the continuous transition along the
line joining A and B in Fig. \ref{phases}, one expects to see the
usual finite size effects.\cite{landau} Most of the earlier
simulations have been carried out in systems of similar sizes. For
example, the system size $L=4$ was also used by Bresme {\it et
al.}\cite{bresme} for a CLG in fcc lattice. Bellaiche and
Vanderbilt\cite{van} chose $L=6$ for their study of cubic perovskite
alloys. For a CLG in sc lattice, larger size lattices have been used
because the number of atoms per unit cell is one in this case. For
example, Panagiotopoulos and Kumar\cite{pana} and Dickman and
Stell\cite{stell} chose $L=12$ and $16$, respectively.

\section{\label{sec:4}Summary}
In summary, our phase diagram has shown the distinct feature of
having two tricritical points for a CLG in fcc lattice. We have
demonstrated that Monte Carlo simulation using an ionic model of
(AgSbTe$_{2}$)$_{x}$(PbTe)$_{2(1-x)}$ shows different possible
microstructural orderings. We have found that layered structures
formed out of AgSbTe$_{2}$ layers separated by Pb$_{2}$Te$_{2}$
layers are generic low temperature structures. In addition to the
layered structures, we have also discovered tubular structures for
$x=0.5$. For $x=1$, a bct structure has been found, in agreement
with previous simulation results. Structures for other values of $x$
are mixtures of those for $x=0$, $0.5$, and $1$. These results will
be discussed in a separate paper.\cite{hoang}

\begin{acknowledgments}
This work is supported by ONR-MURI Program (Contract No.
N00014-02-1-0867). We acknowledge helpful discussions with Professor
M. G. Kanatzidis.
\end{acknowledgments}

%\bibliography{.bib}

\end{document}